\documentclass[aps,prl,showpacs,twocolumn,superscriptaddress]{revtex4}
\usepackage{graphicx,color}
\usepackage{bm}
\usepackage{amsmath}

\def\be{\begin{equation}}
\def\ee{\end{equation}}
\def\bea{\begin{eqnarray}}
\def\eea{\end{eqnarray}}

\begin{document}
\title
{Enhanced spin Hall effect by resonant skew scattering in orbital-selective Kondo effect}
\author{Guang-Yu Guo}
\affiliation{Department of Physics, National Taiwan University, Taipei 106, Taiwan}
\author{Sadamichi Maekawa}
\affiliation{Institute for Materials Research, Tohoku University, Sendai 980-8577, Japan}
\affiliation{CREST, Japan Science and Technology Agency, Tokyo 100-0075, Japan}
\author{Naoto Nagaosa}
\email{nagaosa@appi.t.u-tokyo.ac.jp}
\affiliation{Department of Applied Physics, The University of Tokyo,Tokyo 113-8656, Japan}
\affiliation{Cross-Correlated Materials Research Group (CMRG), ASI, RIKEN, Wako 351-0198, Japan}
\begin{abstract}
The enhanced spin Hall effect in Au metal due to the resonant
skew scattering is studied with first-principles band structure 
calculations. Especially the gigantic
spin Hall angle $\gamma_S \cong 0.1$ observed recently (T.Seki et al.,
Nature Materials {\bf 7}, 125 (2008)) is
attributed to the orbital-dependent Kondo effect of 
Fe in the Au host metal, 
where the $t_{2g}$-orbitals are in the mixed-valence region 
while $e_g$-orbitals are in the Kondo limit. 
The enhanced spin-orbit interaction by the electron correlation in the 
$t_{2g}$-orbitals leads to the gigantic spin Hall effect. 
Impurities with 5d orbitals are also discussed. 
\end{abstract}
\pacs{71.70.Ej, 75.30.Kz, 75.80.+q, 77.80.-e}

\maketitle

Spin Hall effect (SHE) is an effect where the transverse spin current is
produced by the electric field or electric current
\cite{1,2,3,4,5,6,7,8,9,10,11,12,13,14,15}. It does
not require any magnetic field or magnets, and is an interesting
and promising phenomenon for application to spin injection and
manipulation in spintronics.
SHE is especially robust and large in metals due to the large
number of carriers and Fermi energy \cite{10,11,12,13,14,15}.
For example, SHE in Pt metal even at room temperature \cite{12,14}
is more than 2 orders of magnitude larger than
that of GaAs \cite{7}.
Even larger SHE has been recently reported in Au/FePt
System \cite{13}, where the spin Hall conductivity is
$\sim 10^5$ $ \Omega^{-1}$cm$^{-1}$ and the spin Hall angle
$\gamma_S$ is as large as $\cong 0.1$.

Naively speaking, SHE is the two copies of 
anomalous Hall effect (AHE) for up and down spins,
respectively, and the knowledge on the latter can be directly
transferred to the former.
When the longitudinal conductivity $\sigma_{xx}$ is larger than
$\sim 10^6$  $\Omega^{-1} $cm$^{-1}$, and the Hall conductivity 
$\sigma_H$ is much larger than
$e^2/(ha) \cong 10^3$ $\Omega^{-1}$cm$^{-1}$  with $h$ and $a$
being the Planck constant and the lattice constant, respectively,
the dominant contribution to the AHE is the extrinsic skew scattering
\cite{17,22,27,24,25}.
In this case, $\sigma_H$ is proportional to
$\sigma_{xx}$, and hence the Hall angle
$\gamma=\sigma_H/\sigma_{xx}$ is
the well-defined measure for the magnitude of AHE independent of
the impurity concentration. In other words, the Hall angle can be
determined by examining the single scattering event due to the impurity.
The typical value of $\gamma$ is estimated as the ratio
$\sim \lambda /\varepsilon _F$ with $\lambda$ being the spin-orbit
interaction (SOI) and $\varepsilon_F$ the Fermi energy.
Usually, $\varepsilon _F$ is at least a few eV, the $\lambda $ is of
the order of 10-20 meV for the $3d$ orbitals, for example. Therefore,
the Hall angle is roughly estimated as $\gamma \sim 10^{-3}$.
This Hall angle $\gamma$ can be enhanced by the resonant skew scattering
by the magnetic impurity \cite{22,27}. Using the Anderson Hamiltonian
describing the virtual bound state causing the resonant scattering,
the anomalous Hall angle can be expressed in terms of the hybridization
energy $\Delta $ and SOI $\lambda $, and the phase shift $\delta_1$
due to the
$p$-wave scattering as $\gamma \sim (\lambda /\Delta) \sin \delta _1$,
which can be of the order of $10^{-2}$ \cite{22}. Therefore, the spin
Hall angle $\gamma_S \cong 0.1$ is a surprisingly
large value, which needs to be understood for the design of the gigantic
SHE.

 In this paper, we propose that the local electron correlation
and spin fluctuation further enhance SHE compared with AHE 
by the explicit first-principles band
structure calculation with the Au metal as the host where the 
Fermi energy is at the 6$s$ bands. 
Namely, SHE is not the simple two copies of AHE,
and the comparison between these two offers a unique opportunity to study the many-body effect.

\begin{figure}
\includegraphics[width=7cm]{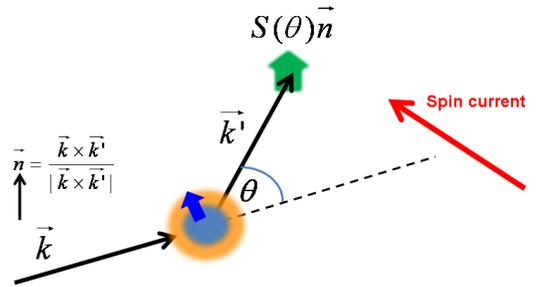}
\caption{\label{Fig1} (color)
The skew scattering due to the spin-orbit interaction
of the scatterer and the spin unpolarized electron beam with wavevector 
${\vec k}$ with the angle $\theta$ with the spin polarization 
$S(\theta) {\vec n}$ with
${\vec n}=({\vec k} \times {\vec k}')/|{\vec k} \times {\vec k}'|$.
}
\end{figure}

   Let us consider the scattering of a spin $S=1/2$ particle by a potential
with SOI. The amplitude of the scattered wave is
given by \cite{17}
\begin{eqnarray}
f_\uparrow (\theta )&=&f_1(\theta )|\uparrow \rangle
+ie^{i\varphi }f_2(\theta )|\downarrow \rangle \nonumber \\
f_\downarrow (\theta )&=&f_1(\theta )|\downarrow \rangle
-ie^{-i\varphi }f_2(\theta )|\uparrow \rangle
\end{eqnarray}
for incoming up-spin and down-spin electrons, respectively, 
where $\theta$ is
the angle between the wavevectors of incident (${\vec k}$) and scattered
($\vec k'$) waves, and $f_1(\theta))(f_2(\theta))$ corresponds to the
spin
non-flip (spin flip) scattering amplitudes. The skewness of the
scattering
is represented by the function
\begin{eqnarray}
S(\theta )=\frac{2{\rm Im}[f^*_1(\theta )f_2(\theta )]}
{|f_1(\theta )|^2+|f_2(\theta )|^2},
\end{eqnarray}
while the function
$I(\theta)=|f_1(\theta)|^2+ f_2(\theta)|^2$ determines the scattering
intensity.
The meaning of the skewness function $S(\theta)$ is shown in Fig.1.
When the unpolarized electrons are incident, the scattered electrons are
spin polarized as $S(\theta) {\vec n}$ with
${\vec n}=({\vec k} \times {\vec k}')/|{\vec k} \times {\vec k}'|$.
By solving the Boltzmann equation with this skew scattering, the spin
Hall angle $\gamma_S$ is given by \cite{9}

\begin{eqnarray}
\gamma_S =\frac{\int d\Omega I(\theta )S(\theta )\sin \theta }
{\int d\Omega I(\theta )(1-\cos \theta )},
\end{eqnarray}
where $\int d \Omega$ is the integral over the solid angle.
The numerator in eq.(3) represents the transverse spin current  
produced by the scattering, i.e., the velocity perpendicular both to
${\vec k}$ and ${\vec n}$ with the spin polarized along ${\vec n}$,
while
the denominator corresponds to the usual transport scattering rate.
Without the resonance effect, the typical value of $\gamma$ is of the
order
of $10^{-3}$, i.e., much smaller than unity. Actually, the obtained
value of
$\gamma$ in ref.\cite{9} is 1/500 for n-type GaAs assuming the screened
Coulomb potential.
The partial wave analysis gives the expression for $f_1(\theta)$ and
$f_2(\theta)$
\begin{eqnarray}
f_1(\theta )&=&\sum _l \frac{P_l(\cos \theta )}{2ik}\left
[
(l+1)\left (e^{2i\delta ^+_l}-1\right )+l
\left (e^{-2i\delta ^-_l}-1\right )\right ],
\nonumber \\
f_2(\theta )&=&\sum _l\frac{\sin \theta }{2ik}\left
(
e^{2i\delta ^+_l}-e^{2i\delta ^-_l}\right )
\frac{d}{d\cos \theta } P_l(\cos \theta ) .
\end{eqnarray}
Putting eq.(4) into eq.(3),  we obtain 
\begin{equation}
\gamma_S =\frac{3}{2}\frac{{\rm Im}
\left [\left (e^{-2i\delta _1}-1\right )
\left (e^{2i\delta ^+_2}-e^{2i\delta ^-_2}\right )\right ]}
{9\sin ^2\delta _2^++4\sin ^2\delta _2^-
+3 \left[1-\cos 2\left (\delta ^+_2-\delta ^-_2 \right) \right] }
\end{equation}
where we have assumed that the resonant channel is the $d$-wave ($l=2$)
which is subject to SOI and the scattering is
characterized by the two phase shifts $\delta^{\pm}_2 =
\delta_{J = 2 \pm 1/2}$, while that ($\delta_1$) for the $p$-wave
scattering
is assumed to be spin-independent. Assuming that $\delta_1$ for the
non-resonant $p$-wave scattering is small
( $|\delta_1| \cong 0.1$ \cite{22}), $\gamma_S \cong -3 \delta_1
(\cos 2 \delta^+_2 - \cos 2 \delta^-_2) /
(9\sin ^2\delta _2^++4\sin ^2\delta _2^-
+3\left[1-\cos 2(\delta ^+_2-\delta ^-_2) \right] )$.
Therefore, the most
important factor is $ \cos 2 \delta^+_2 - \cos 2 \delta^-_2 $, which
is related to the difference in the occupation numbers of the
$J= 2 \pm 1/2$ impurity states induced by SOI
through the Friedel sum rule \cite{26}. Therefore, the local density of states
(DOS) for the $d$-electrons determines the magnitude of SHE, which can be 
studied by the first-principles band structure calculation as described below.

\begin{figure}[h]
\includegraphics[width=7cm]{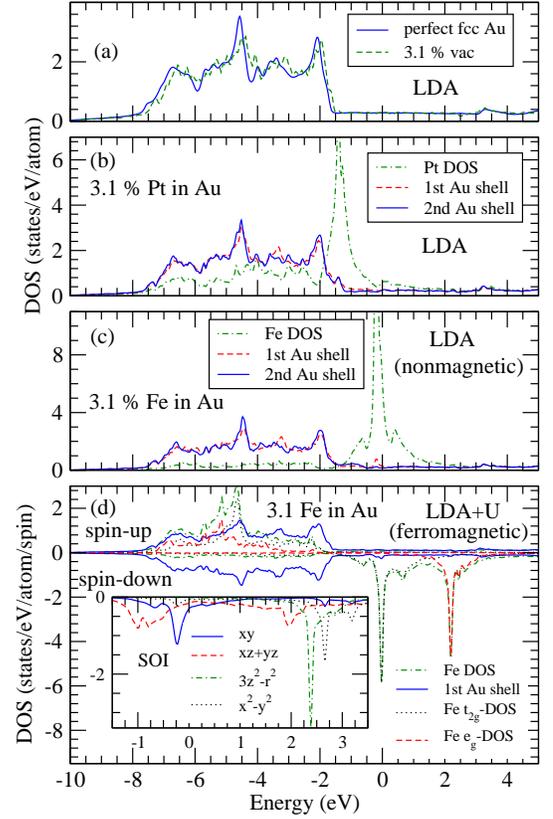}
\caption{\label{Fig2} (color)
Total, site-, orbital- and spin-decomposed DOSs
of (a) bulk fcc Au, 3.1 \% Au-vacancies, (b) 3.1 \% Pt-impurities,
(c) 3.1 \% Fe-impurities in nonmagnetic state and (d) in ferromagnetic
state.}
\end{figure}

We take Au as the host metal, where the gigantic SHE has been
observed \cite{13}. The electron configuration of Au atom is
$5d^{10} 6s^1$, and the Fermi energy is at the 6$s$6$p$-bands
with relatively small SOI.
Therefore, it is difficult to explain the giant SHE in terms of the
intrinsic mechanism, in sharp contrast to the case of Pt with $5d^{9} 6s^2$ 
\cite{14}.
Because in \cite{13} the spin current is supplied from FePt to Au,
it is natural to consider the three possibilities of the imperfections,
i.e., (i) Au vacancies, (ii) Pt impurities, and (iii) Fe impurities.
From the above consideration, there are several requirements to obtain
the gigantic SHE;
(I) there should be the resonance at the Fermi energy due to either the
Kondo peak or the mixed valance, (II) the orbital angular momentum should
not be quenched, and (III) the peak must be split due to SOI
 by the energy comparable or larger than the width of the peak.
We have studied the impurity states in Au host metal 
for these three cases in terms of the local density approximation 
(LDA)\cite{vos80} plus on-site Coulomb
interaction $U$ (LDA+$U$) \cite{lie95} ($U = 5, J = 0.9 eV$).
We used the highly accurate full-potential augmented plane wave method,
as implemented in the WIEN2k code \cite{wien2k}. 
The atomic positions are relaxed in the presence of the impurity.
The numerical results presented below have been tested for convergence
with respect to the energy cut-off for augmented plane waves, $k$-points
used for irreducible Brillouin zone and supercell size. 
Figure 2 shows the density of states (DOS) in the presence of one impurity in the
$2\times2\times2$ supercell. 
Basically the structure extending from -8 eV to -2 eV is due to 
the 5$d$ bands of
Au, while 6$s6p$ bands of Au are extended all through in this energy
range
with smaller DOS. The change in the DOS is almost confined in the range
of
the 5$d$ bands in the case of (i) vacancy (Fig.2(a)), and is at around -1.5
eV in
the case of (ii) Pt (Fig.2(b)). We could not obtain a magnetic state
in both
of these cases even in the LDA+$U$ calculation. This conclusion is
consistent
with the earlier study by photoelectron spectroscopy \cite{28}.
In sharp contrast, we obtained the magnetic state as the 
ground state with both the LDA and LDA+$U$
calculations in the case of Fe. Accordingly, the DOS for Fe
(Fig.2(d)) shows the spin splitting, and the down spin 
DOS has the sharp peak close to the Fermi energy. 
(We put the DOS for nonmagnetic state in Fig.2(c) for reference.)
This means that there occurs the valence fluctuation of Fe ion between
$d^6$ and $d^7$. This is reasonable since the Fe in Au is known as a
Kondo impurity but the Kondo temperature is as low as 0.4 K. From these
results, we conclude that only (iii) Fe impurities could the origin of 
the large SHE satisfying the criteria (I),(II),(III) discussed above, 
and we now analyze Fe impurity in Au in more detail below.

\begin{table}
\caption{Down-spin occupation numbers of the 3$d$-orbitals of the Fe impurity
in Au from LDA+$U$ calculations without SOI and with SOI. The calculated 
magnetic moments are: $m_s^{Fe}$ = 3.39 $\mu_B$ and $m_s^{tot}$ = 3.32 $\mu_B$
without SOI, as well as $m_s^{Fe}$ = 3.19 $\mu_B$, $m_o^{Fe}$ = 1.54 $\mu_B$ 
and $m_s^{tot}$ = 3.27 $\mu_B$ 
with SOI. The muffin-tin sphere radius $R_{mt} = 2.65 a_0$ ($a_0$ is Bohr radius)
is used for both Fe and Au atoms. }
\begin{ruledtabular}
\begin{tabular}{cccccc}

(a) & $xy$ & $xz$ & $yz$ & $3z^2-r^2$ & $x^2-y^2$ \\
no SOI & 0.459 & 0.459 & 0.459 & 0.053 & 0.053    \\
SOI  & 0.559& 0.453 & 0.453  & 0.050 & 0.128  \\ \hline

(b) & $m =-2$ & $m =-1$   & $m = 0$   & $m = 1$ & $m = 2$  \\
no SOI & 0.256 &0.459 & 0.053  & 0.459& 0.256  \\
SOI & 0.138 & 0.087 & 0.050 & 0.819& 0.549  \\

\end{tabular}
\end{ruledtabular}
\end{table}

The band structure calculation for the state with local magnetic and orbital
orderings corresponds to the mean field theory of the Anderson
Hamiltonian \cite{26}. 
In the limit of isolated impurity atom, this mean field 
theory gives the correct energy positions of the peaks in the  
spectral function of the single-particle Green's function.
Of course, there is no symmetry breaking by the local electron correlation,
and the true ground state is the quantum mechanical superposition of
the degenerate symmetry breaking states, i.e., quantum fluctuation of 
spins and orbitals occur.   
Early theories of Kondo effect of Fe in Au have estimated the relevant
quantities as $\Delta \cong 1.4$eV, $U \cong 5.4$eV, $J \cong
0.9$eV \cite{29}, where $\Delta$ is the energy broadening of the virtual 
bound state due to the hybridization with the conduction bands, 
$U$ is the on-site Coulomb interaction, and $J$ is the Hund's coupling
energy between different orbitals. The crystal field splitting is
considered to be small ($\sim 0.1$eV), and SOI $\lambda \cong 30$meV is even
smaller. The resistivity measurement at room temperature shows
a systematic change as the valence of the 
impurity changes as Ti, V, Cr, Mn, Fe, Co, Ni, and shows the dip at Mn, 
while the maximum at Fe (Fig. 17 of
Ref. \onlinecite{30}). This strongly suggests the peak in the DOS for Fe,
i.e., the mixed
valence case, although the low temperature properties have been analyzed by
the Kondo model \cite{30}. The magnetic susceptibility measurement shows
the
$S=2$ magnetic moment, while the Moessbauer experiment concluded rather
$S=3/2$ \cite{31}.  
These somewhat confusing situation can be resolved by the LDOS in
Fig. 2(c-d) together with Table I. 
Table I shows the spin/orbital magnetic moments and the 
occupation number of each $d$-orbital in terms of LDA + SOI + $U$ 
calculation. Note that the obtained values depend slightly
on the muffin-tin sphere radius $R_{mt}$ and $U$ value, 
but the semi-quantitative conclusion does not change.  
First, the nonmagnetic state in the LDA calculation shows almost
no crystal field splitting (of the order of 0.1eV), which is consistent
with the earlier result. However, the inclusion of $U$ changes the situation
dramatically, and the $e_g-t_{2g}$ splitting is enhanced to be around
2 eV as shown in Fig. 2(d). This corresponds to the orbital polarization
due to the electron correlation, i.e., the local version of the orbital
ordering not by the crystal field but by the electron correlation 
\cite{33}.
Therefore we conclude that the orbital-dependent Kondo effect occurs for
Fe in Au; the $e_g$ orbitals are in the Kondo limit, while 
$t_{2g}$-orbitals are in the mixed valence region. 
Therefore, at temperatures above $T_K \cong
0.4$ K, the $t_{2g}$-orbitals, within which the orbital angular momentum
is not quenched, play the major role in the transport properties, while the
$e_g$-orbitals determine the low temperature Kondo effect. 

Now we consider the orbital polarization within the $t_{2g}$ states due to the 
SOI. Note that $t_{2g}$-orbitals behaves like $l_{\rm eff.} =1$
states with $xy$,$zx+iyz$, and $zx-iyz$ orbitals corresponding to 
$m=0$, $m=1$, and $m=-1$ states, respectively. Therefore, SOI is 
effective within the $t_{2g}$ states leading to the energy 
splitting between the $J_{\rm eff.} = 3/2$ and $J_{\rm eff.} = 1/2$ states.
Naively, the orbital polarization is determined by the competition 
between the hybridization energy $\Delta$ and the energy splitting 
due to SOI. Here, one must carefully distinguish between the many-body states and
the single-particle states. The energy separation between the many-body state with different total angular momentum $J$ is typically the order of SOI, and much smaller 
than $U$ or $J$. However, once the many-body ground state of $d$-electrons is 
fixed, the separation of the single-particle state energy, which is the 
energy difference between the $N$-electron and $N \pm1$-electron state,
can be as large as $U$ or $J$, which is important for the conduction 
electrons which comes in or out to the $d$-orbitals.
Therefore, it is possible that the electron correlation $U$ plays an 
essential role, and even a SOI much smaller than the hybridization energy
can produce the large orbital magnetic moment 
$m_o = 1.54 \mu_B$ as shown in Table I. 
Correspondingly, the single-particle 
$m=1$ state is almost occupied (0.819) while $m=-1$ state is almost empty 
(0.087), as is seen also from the inset of Fig. 2(d).
Figure 2(d) and Table I also explain why the AHE is rather small compared
with SHE. 
As for AHE, the phase shift for $m=1$ ($m=-1$) are almost 
$\pi$ ($0$), and both of them does contribute a little to the scattering, while
$m=0$ state is almost in the unitary limit with $\pi/2$-phase shift.
Therefore, we do not expect the enhanced AHE, 
leading to the small anomalous Hall angle 
compared with that for SHE \cite{22}. 
For the SHE, on the other hand, such a cancelation does not occur.
One needs to treat the spin/orbital fluctuation in this case, 
but a rough estimate for the spin Hall angle can be obtained as follows.
For the conduction electrons, the energy difference 
between $J_{\rm eff}=3/2$ and  $J_{\rm eff}=1/2$ is that of 
$m=1$ and $m=-1$ states in the mean field theory when one 
considers the Ising type coupling $l_z s_z$. This energy 
is larger than the hybridization energy and 
we expect the difference of the phase shifts between these two 
channels of the order of $\pi$, giving a large spin Hall angle of the 
order of $\delta_1 \cong 0.1$ as observed experimentally \cite{13}.
This picture for the Kondo effect in Fe impurity in Au is 
different from the conventional one \cite{29}.
Further experimental studies using the spectroscopies such as STS and
ARPES are highly desired to clarify the nature of this fundamental 
problem.

\begin{figure}[h]
\includegraphics[width=7cm]{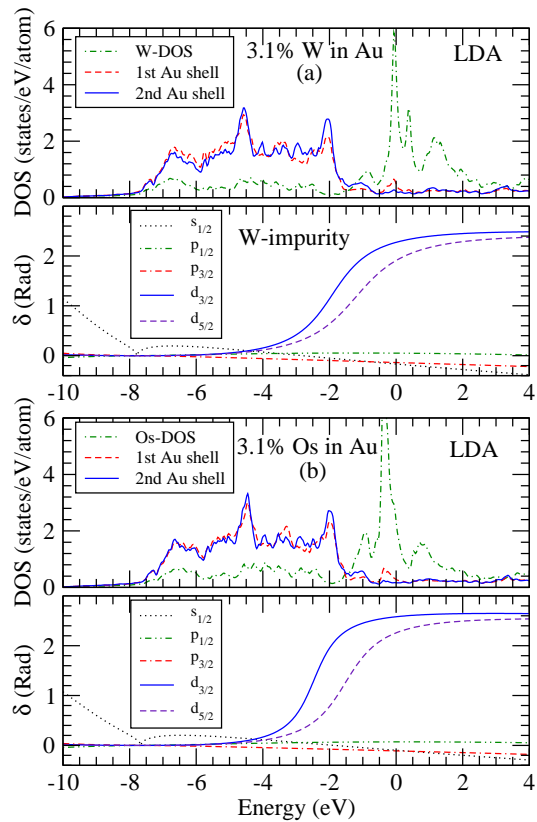}
\caption{\label{Fig3} (color)
Site-decomposed DOSs and phase shifts ($\delta$)
of (a) 3.1 \% W and (b) 3.1 \% Os in Au.}
\end{figure}

SHE associated with Kondo effect is expected for other ions such as rare-earth
impurity \cite{Kontani}.  From this viewpoint, we have also studied the 
case of 5$d$ impurities, i.e., Ta, W, Re, Os, Ir doped into Au. 
In these cases, the appropriate $U$ value is around 0.5 eV, and there
occurs no magnetic state. However, SOI is almost as large as $U$, and
hence the large SHE is expected. 
The results for the DOS and the associated phase
shifts are shown in Fig.3  for W (Fig.3(a)) and Os (Fig.3(b)) as the two
representative examples. Even without the magnetic
moment, SHE is enhanced due to
the resonant state at Fermi energy, and the corresponding spin Hall
angle is 
estimated as $\gamma \cong 0.2 \sin 2 \delta_1$ where $\delta_1$ is the
phase shift for the $p$-wave scattering, and we expect
$|\gamma_S| \cong 0.04$ in these cases in semi-quantitatively agreement
with the early experiment \cite{22}.

In conclusion, we have studied the enhanced SHE due to the resonant skew
scattering. 
Compared with AHE, SHE, where the time-reversal symmetry is not broken, 
gives a unique opportunity to study the electronic states near the
impurity without disturbing its magnetic behavior. We have presented the formula for
the spin Hall angle in terms of the phase shifts (eq.(5)), and also the
first-principles band structure calculation to analyze the mechanism of
the gigantic SHE. We have shown that SHE shed a new light on the 
Kondo effect, which plays the key role to enhance SHE and 
a new picture for the Fe impurity state in Au has been proposed. 
This leads to the material design of the large SHE even at room
temperature with the potential application to the spintronics.

We thank Profs. A. Fert, K.Takanashi, N. Kawakami and Y. Otani for
fruitful discussions.

\end{document}